\documentstyle[12pt]{article}
\begin{document}

\thispagestyle{empty}
\renewcommand{\thefootnote}{\fnsymbol{footnote}}
\setcounter{footnote}{0}

\def\pxb{\left(\p \times \B - \B \times \p \right)}
\def\LAMBDA{\bar{\lambda}}
\def\rk{r_k}
\def\beq{\begin{equation}}
\def\eeq{\end{equation}}
\def\bea{\begin{eqnarray}}
\def\eea{\end{eqnarray}}
\def\nn{\nonumber}
\def\ba{\begin{array}}
\def\ea{\end{array}}
\def\0{{\mbox{\boldmath $0$}}}
\def\one{1\hskip -1mm{\rm l}}
\def\A{{\mbox{\boldmath $A$}}}
\def\B{{\mbox{\boldmath $B$}}}
\def\El{{\mbox{\boldmath $E$}}}
\def\F{{\mbox{\boldmath $F$}}}
\def\S{{\mbox{\boldmath $S$}}}

\def\a{{\mbox{\boldmath $a$}}}
\def\p{{\mbox{\boldmath $p$}}}
\def\hatp{{\hat{\mbox{\boldmath $p$}}}}
\def\hatP{{\hat{\mbox{\boldmath $P$}}}}
\def\vpi{{\mbox{\boldmath $\pi$}}}
\def\hatvpi{\hat{\mbox{\boldmath $\pi$}}}
\def\r{{\mbox{\boldmath $r$}}}
\def\v{{\mbox{\boldmath $v$}}}
\def\w{{\mbox{\boldmath $w$}}}
\def\H{{\rm H}}
\def\hA{\hat{A}}
\def\hB{\hat{B}}
\def\i{{\rm i}}
\def\ih{\frac{\i}{\hbar}}
\def\ixh{\i \hbar}
\def\ddx{\frac{\partial}{\partial x}}
\def\ddy{\frac{\partial}{\partial y}}
\def\ddz{\frac{\partial}{\partial z}}
\def\ddt{\frac{\partial}{\partial t}}
\def\vsig{{\mbox{\boldmath $\sigma$}}}
\def\Al{{\mbox{\boldmath $\alpha$}}}
\def\ho{\hat{\cal H}_o}
\def\half{\frac{1}{2}}
\def\E{{\hat{\cal E}}}
\def\O{{\hat{\cal O}}}
\def\eps{\epsilon}
\def\g{\gamma}
\def\Vomeg{{\underline{\mbox{\boldmath $\Omega$}}}_s}
\def\hH{\hat{H}}
\def\Vsig{{\mbox{\boldmath $\Sigma$}}}
\def\Nab{{\mbox{\boldmath $\nabla$}}}
\def\curl{{\rm curl}}
\def\bh{\bar{H}}
\def\th{\tilde{H}}

\def\zone{z^{(1)}}
\def\ztwo{z^{(2)}}
\def\zi{z_{\rm in}}
\def\zo{z_{\rm out}}

\def\At{{\hat{A}(t)}}
\def\dAt{\frac{\partial {\hat{A}(t)} }{\partial t}}
\def\sone{{\hat{S}_1}}
\def\dsone{\frac{\partial {\hat{S}_1} }{\partial t}}
\def\dO{\frac{\partial {\hat{\cal O}}}{\partial t}}

\begin{center}
{\Large\bf
Maxwell Optics: II. An Exact Formalism}

\bigskip

{\em Sameen Ahmed KHAN} \\

\bigskip

khan@fis.unam.mx ~~~ http://www.pd.infn.it/$\sim$khan/ \\
rohelakhan@yahoo.com ~~~ http://www.imsc.ernet.in/~jagan/khan-cv.html \\
Centro de Ciencias F\'{i}sicas,
Universidad Nacional Aut\'onoma de M\'exico, \\
Apartado Postal 48-3,
Cuernavaca 62251,
Morelos, \\
{\bf M\'EXICO} \\

\end{center}

\bigskip

\noindent
{\bf Abstract} \\
We present a formalism for light optics starting with the Maxwell
equations and casting them into an exact matrix form taking into
account the spatial and temporal variations of the permittivity and
permeability.  This $8 \times 8$ matrix representation is used to
construct the optical Hamiltonian.  This has a close analogy with
the algebraic structure of the Dirac equation, enabling the use of
the rich machinery of the Dirac electron theory.  We get interesting
wavelength-dependent contributions which can not be obtained in any
of the traditional approaches.

\bigskip

\section{Introduction}
The traditional scalar wave theory of optics (including aberrations to
all orders) is based on the beam-optical Hamiltonian derived using the
Fermat's principle.  This approach is purely geometrical and works
adequately in the scalar regime.  The other approach is based on the
Helmholtz equation which is derived from the Maxwell equations.  Then
one makes the {\em square-root} of the Helmholtz operator followed by
an expansion of the radical~\cite{DFW,Dragt-Wave}.  This approach
works to all orders and the resulting expansion is no different from
the one obtained using the geometrical approach of the Fermat's
principle.

Another way of obtaining the aberration expansion is based on the
algebraic similarities between the Helmholtz equation and the
Klein-Gordon equation.  Exploiting this algebraic similarity the
Helmholtz equation is linearized in a procedure very similar to the one
due to Feschbach-Villars, for linearizing the  Klein-Gordon equation.
This brings the Helmholtz equation to a Dirac-like form and then
follows the procedure of the Foldy-Wouthuysen expansion used in the
Dirac electron theory.  This  approach, which uses the algebraic
machinery of quantum mechanics, was developed recently~\cite{KJS-1},
providing an alternative to the traditional {\em square-root}
procedure.  This scalar formalism gives rise to wavelength-dependent
contributions modifying the aberration coefficients~\cite{Khan-1}.
The algebraic machinery of this formalism is very similar to the one
used in the {\em quantum theory of charged-particle beam optics}, based
on the Dirac~\cite{JSSM} and the Klein-Gordon~\cite{KJ1} equations
respectively.  The detailed account for both of these is available
in~\cite{JK2}.  A treatment of beam optics taking into account the
anomalous magnetic moment is available in~\cite{CJKP-1}.

As for the polarization: A systematic procedure for the passage from
scalar to vector wave optics to handle paraxial beam propagation
problems, completely taking into account the way in which the Maxwell
equations couple the spatial variation and polarization of light
waves, has been formulated by analysing the basic Poincar\'{e}
invariance of the system, and this procedure has been successfully
used to clarify several issues in Maxwell
optics~~\cite{MSS-1,SSM-1,SSM-2}.

In all the above approaches, the beam-optics and the polarization are
studied separately, using very different machineries.
The derivation of the Helmholtz equation from the Maxwell equations is
an approximation as one neglects the spatial and temporal derivatives
of the permittivity and permeability of the medium.  Any prescription
based on the Helmholtz equation is bound to be an approximation,
irrespective of how good it may be in certain situations.  It is very
natural to look for a prescription based fully on the Maxwell equations.
Such a prescription is sure to provide a deeper understanding of
beam-optics and polarization in a unified manner.  With this as the
chief motivation we construct a formalism starting with the Maxwell
equations in a matrix form: a single entity containing all the four
Maxwell equations.

In our approach we require an exact matrix representation of the
Maxwell equations in a medium taking into account the spatial and
temporal variations of the permittivity and permeability.  It is
necessary and sufficient to use $8 \times 8$ matrices for such an exact
representation.  The derivation of the required matrix representation,
and how it differs from the numerous other ones is presented in
Part-I~\cite{Khan-2}.

In the present Part (Part-II) we proceed with the exact matrix
representation of the Maxwell equations derived in Part-I, and
construct a general formalism.  The derived representation has a very
close algebraic correspondence with the Dirac equation.  This enables
us to apply the machinery of the Foldy-Wouthuysen expansion used in the
Dirac electron theory.  The Foldy-Wouthuysen transformation technique
is outlined in Appendix-A.  General expressions for the Hamiltonians
are derived without assuming any specific form for the refractive
index.  These Hamiltonians are shown to contain the extra
wavelength-dependent contributions which arise very naturally in our
approach.  In Part-III ~\cite{Khan-4} we apply the general formalism to
the specific examples:
A. {\em Medium with Constant Refractive Index}.  This example
is essentially for illustrating some of the details of the machinery
used.

The other application, B. {\em Axially Symmetric Graded Index Medium}
is used to demonstrate the power of the formalism.  Two points are worth
mentioning, {\em Image Rotation}:  Our formalism gives rise to the
image rotation (proportional to the wavelength) and we have derived an
explicit relationship for the angle of the image rotation.  The other
pertains to the aberrations: In our formalism we get all the nine
aberrations permitted by the axial symmetry.  The traditional approaches
give six aberrations.  Our formalism modifies these six aberration
coefficients by wavelength-dependent contributions and also gives rise
to the remaining three permitted by the axial symmetry.  The existence
of the nine aberrations and image rotation are well-known in {\em
axially symmetric magnetic lenses}, even when treated classically.  The
quantum treatment of the same system leads to the wavelength-dependent
modifications~\cite{JK2}.  The alternate procedure for the Helmholtz
optics in~\cite{KJS-1, Khan-1} gives the usual six aberrations
(though modified by the wavelength-dependent contributions) and does
not give any image rotation.  These extra aberrations and the image
rotation are the exclusive outcome of the fact that the formalism
is based on the Maxwell equations, and done exactly.

The traditional beam-optics is completely obtained from our approach
in the limit wavelength, $\LAMBDA \longrightarrow 0$, which we call as
the traditional limit of our formalism.  This is analogous to the
classical limit obtained by taking $\hbar \longrightarrow 0$ in the
quantum prescriptions.  The scheme of using the Foldy-Wouthuysen
machinery in this formalism is very similar to the one used in the
{\em quantum theory of charged-particle beam
optics}~\cite{JSSM,KJ1,JK2}.  There  too one recovers the classical
prescriptions in the limit $\lambda_0 \longrightarrow 0$ where
$\lambda_0 = {\hbar}/{p_0}$ is the de Broglie wavelength and $p_0$ is
the design momentum of the system under study.

The studies on the polarization are in progress.  Some of the results
in~\cite{SSM-2} have been obtained as the lowest order approximation
of the more general framework presented here.  These will be presented
in Part-IV soon~\cite{Khan-5}.

\section{An exact matrix representation of the \\
Maxwell equations in a medium}
Matrix representations of the Maxwell equations are very
well-known~\cite{Moses}-\cite{Birula}.  However, all these
representations lack an exactness or/and are given in terms of a
{\em pair} of matrix equations.
A treatment expressing the Maxwell equations in a single matrix
equation instead of a {\em pair} of matrix equations was obtained
recently~\cite{Khan-2}.  This representation contains all the four
Maxwell equations in presence of sources taking into account the
spatial and temporal variations of the permittivity
$\epsilon (\r , t)$ and the permeability $\mu (\r , t)$.

Maxwell equations~\cite{Jackson, Ponofsky-Phillips} in an inhomogeneous
medium with sources are
\bea
\Nab \cdot {\mbox{\boldmath $D$}} \left(\r , t \right)
=
\rho\,, \nn \\
\Nab \times {\mbox{\boldmath $H$}} \left(\r , t \right)
- \frac{\partial }{\partial t}
{\mbox{\boldmath $D$}} \left(\r , t \right)
=
{\mbox{\boldmath $J$}}\,, \nn \\
\Nab \times {\mbox{\boldmath $E$}} \left(\r , t \right)
+
\frac{\partial }{\partial t}
{\mbox{\boldmath $B$}} \left(\r , t \right)
= 0\,, \nn \\
\Nab \cdot {\mbox{\boldmath $B$}} \left(\r , t \right)
= 0\,.
\label{Maxwell-1}
\eea
We assume the media to be linear, that is
${\mbox{\boldmath $D$}} = \epsilon (\r , t) \El$, and
${\mbox{\boldmath $B$}} = \mu (\r , t) {\mbox{\boldmath $H$}}$,
where $\epsilon$ is the {\bf permittivity of the medium} and
$\mu$ is the {\bf permeability of the medium}.
The magnitude of the velocity of light in the medium is given by
$v (\r , t)  = \left|{\mbox{\boldmath $v$}} (\r , t)\right|
= {1}/{\sqrt{\epsilon (\r , t) \mu (\r , t)}}$.  In vacuum we
have, $\epsilon_0 = 8.85 \times 10^{- 12} {C^2}/{N. m^2}$ and
$\mu_0 = 4 \pi \times 10^{- 7} {N}/{A^2}$.
Following the notation in~\cite{Birula, Khan-2} we use the
Riemann-Silberstein vector given by
\bea
\F^{\pm} \left(\r , t \right)
& = &
\frac{1}{\sqrt{2}}
\left(
\sqrt{\epsilon (\r , t)} \El \left(\r , t \right)
\pm \i \frac{1}{\sqrt{\mu (\r , t)}} \B \left(\r , t \right) \right)\,.
\label{R-S-Conjugate}
\eea
We further define,
\bea
\Psi^{\pm} (\r , t)
=
\left[
\ba{c}
- F_x^{+}  \pm \i F_y^{+} \\
F_z^{+} \\
F_z^{+} \\
F_x^{+} \pm \i F_y^{+}
\ea
\right]\,, \quad
W^{\pm}
=
\left(\frac{1}{\sqrt{2 \epsilon}}\right)
\left[
\ba{c}
- J_x \pm \i J_y \\
J_z - v \rho \\
J_z + v \rho \\
J_x \pm \i J_y
\ea
\right]\,,
\eea
where $W^{\pm}$ are the vectors for the sources.  Following the
notation in~\cite{Khan-2} the exact matrix representation of the
Maxwell equations is
\bea
& &
\frac{\partial }{\partial t}
\left[
\ba{cc}
{\mbox{\boldmath $I$}} & {\mbox{\boldmath $0$}} \\
{\mbox{\boldmath $0$}} & {\mbox{\boldmath $I$}}
\ea
\right]
\left[
\ba{cc}
\Psi^{+} \\
\Psi^{-}
\ea
\right]
-
\frac{\dot{v} (\r , t)}{2 v (\r , t)}
\left[
\ba{cc}
{\mbox{\boldmath $I$}} & {\mbox{\boldmath $0$}} \\
{\mbox{\boldmath $0$}} & {\mbox{\boldmath $I$}}
\ea
\right]
\left[
\ba{cc}
\Psi^{+} \\
\Psi^{-}
\ea
\right] \nn \\
& & \qquad \qquad \quad
+
\frac{\dot{h} (\r , t)}{2 h (\r , t)}
\left[
\ba{cc}
{\mbox{\boldmath $0$}} & \i \beta \alpha_y \\
\i \beta \alpha_y & {\mbox{\boldmath $0$}}
\ea
\right]
\left[
\ba{cc}
\Psi^{+} \\
\Psi^{-}
\ea
\right]
\nn \\
& & \qquad \quad
=  - v (\r , t)
\left[
\ba{ccc}
\left\{
{\mbox{\boldmath $M$}} \cdot \Nab
+
{\mbox{\boldmath $\Sigma$}} \cdot {\mbox{\boldmath $u$}}
\right\}
& &
- \i \beta
\left({\mbox{\boldmath $\Sigma$}} \cdot {\mbox{\boldmath $w$}}\right)
\alpha_y
\\
- \i \beta
\left({\mbox{\boldmath $\Sigma$}}^{*} \cdot {\mbox{\boldmath $w$}}\right)
\alpha_y
& &
\left\{
{\mbox{\boldmath $M$}}^{*} \cdot \Nab
+
{\mbox{\boldmath $\Sigma$}}^{*} \cdot {\mbox{\boldmath $u$}}
\right\}
\ea
\right]
\left[
\ba{cc}
\Psi^{+} \\
\Psi^{-}
\ea
\right] \nn \\
& & \qquad \quad \quad
- \left[
\ba{cc}
{\mbox{\boldmath $I$}} & {\mbox{\boldmath $0$}} \\
{\mbox{\boldmath $0$}} & {\mbox{\boldmath $I$}}
\ea
\right]
\left[
\ba{c}
W^{+} \\
W^{-}
\ea
\right]\,,
\eea
where `$^{*}$' denotes complex-conjugation,
$\dot{v} = \frac{\partial v}{\partial t}$ and
$\dot{h} = \frac{\partial h}{\partial t}$.
The various matrices are
\bea
M_x
& = &
\left[
\ba{cc}
{\mbox{\boldmath $0$}} & \one \\
\one & {\mbox{\boldmath $0$}}
\ea
\right]\,, \qquad
M_y
=
\left[
\ba{cc}
{\mbox{\boldmath $0$}} & - \i \one \\
\i \one & {\mbox{\boldmath $0$}}
\ea
\right]\,, \qquad
M_z
=
\beta
=
\left[
\ba{cc}
\one & {\mbox{\boldmath $0$}} \\
{\mbox{\boldmath $0$}} & - \one
\ea
\right]\,, \nn \\
{\mbox{\boldmath $\Sigma$}}
& = &
\left[
\ba{cc}
{\mbox{\boldmath $\sigma$}} & {\mbox{\boldmath $0$}} \\
{\mbox{\boldmath $0$}} & {\mbox{\boldmath $\sigma$}}
\ea
\right]\,, \qquad
{\mbox{\boldmath $\alpha$}}
=
\left[
\ba{cc}
{\mbox{\boldmath $0$}} & {\mbox{\boldmath $\sigma$}} \\
{\mbox{\boldmath $\sigma$}} & {\mbox{\boldmath $0$}}
\ea
\right]\,, \qquad
{\mbox{\boldmath $I$}}
=
\left[
\ba{cc}
\one & {\mbox{\boldmath $0$}} \\
{\mbox{\boldmath $0$}} & \one
\ea
\right]\,,
\eea
and $\one$ is the $2 \times 2$ unit matrix.  The triplet of the Pauli
matrices, ${\mbox{\boldmath $\sigma$}}$ is
\bea
{\mbox{\boldmath $\sigma$}}
& = &
\left[
\sigma_x =
\left[
\ba{cc}
0 & 1 \\
1 & 0
\ea
\right]\,, \
\sigma_y =
\left[
\ba{lr}
0 & - \i \\
\i & 0
\ea
\right]\,, \
\sigma_z =
\left[
\ba{lr}
1 & 0 \\
0 & -1
\ea
\right]
\right]\,,
\eea
and
\bea
{\mbox{\boldmath $u$}} (\r , t)
& = &
\frac{1}{2 v (\r , t)} \Nab v (\r , t)
=
\frac{1}{2} \Nab \left\{\ln v (\r , t) \right\}
=
- \frac{1}{2} \Nab \left\{\ln n (\r , t) \right\} \nn \\
{\mbox{\boldmath $w$}} (\r , t)
& = &
\frac{1}{2 h (\r , t)} \Nab h (\r , t)
=
\frac{1}{2} \Nab \left\{\ln h (\r , t) \right\}\,.
\label{u-w}
\eea
Lastly,
\bea
{\rm Velocity ~ Function:} \, v (\r , t)
& = &
\frac{1}{\sqrt{\epsilon (\r , t) \mu (\r , t)}} \nn \\
{\rm Resistance ~ Function:} \, h (\r , t)
& = &
\sqrt{\frac{\mu (\r , t)}{\epsilon (\r , t)}}\,.
\eea
As we shall see soon, it is advantageous to use the above derived
functions instead of the permittivity, $\epsilon (\r , t)$ and the
permeability, $\mu (\r , t)$.  The functions, $v (\r , t)$ and
$h (\r , t)$ have the dimensions of velocity and resistance
respectively.

Let us consider the case without any sources~($W^{\pm} = 0$).
We further assume,
\bea
\Psi^{\pm} (\r , t)
=
\psi^{\pm} \left(\r \right) e^{- \i \omega t}\,, \qquad
\omega > 0\,,
\label{Time-Gone}
\eea
with $\dot{v} (\r , t) = 0$ and $\dot{h} (\r , t) = 0$.
Then,
\bea
& &
\left[
\ba{cc}
M_z & {\mbox{\boldmath $0$}} \\
{\mbox{\boldmath $0$}} & M_z
\ea
\right]
\frac{\partial }{\partial z}
\left[
\ba{cc}
\psi^{+} \\
\psi^{-}
\ea
\right] \nn \\
& & \quad
=
\i \frac{\omega}{v (\r)}
\left[
\ba{cc}
\psi^{+} \\
\psi^{-}
\ea
\right] \nn \\
& & \quad \quad
- v (\r)
\left[
\ba{ccc}
\left\{
{\mbox{\boldmath $M$}}_\perp \cdot \Nab_\perp
+
{\mbox{\boldmath $\Sigma$}} \cdot {\mbox{\boldmath $u$}}
\right\}
& &
- \i \beta
\left({\mbox{\boldmath $\Sigma$}} \cdot {\mbox{\boldmath $w$}}\right)
\alpha_y
\\
- \i \beta
\left({\mbox{\boldmath $\Sigma$}}^{*} \cdot {\mbox{\boldmath $w$}}\right)
\alpha_y
& &
- \left\{
{\mbox{\boldmath $M$}}_\perp^{*} \cdot \Nab_\perp
+
{\mbox{\boldmath $\Sigma$}}^{*} \cdot {\mbox{\boldmath $u$}}
\right\}
\ea
\right]
\left[
\ba{cc}
\psi^{+} \\
\psi^{-}
\ea
\right]\,. \nn \\
\label{intermediate}
\eea
At this stage we introduce the process of {\em wavization}, through
the familiar Schr\"{o}dinger replacement
\bea
- \i \LAMBDA \Nab_{\perp} \longrightarrow \hatp_{\perp}\,,
\qquad \qquad
- \i \LAMBDA \frac{\partial }{\partial z}
\longrightarrow p_z\,,
\label{wavization-1}
\eea
where $\LAMBDA = {\lambda}/{2 \pi}$ is the reduced wavelength,
$c = \LAMBDA \omega$ and $n (\r) = {c}/{v (\r)}$ is the refractive
index of the medium.  Noting, that $(pq - qp) = - \i \LAMBDA$, which is
very similar to the commutation relation, $(pq - qp) = - \i \hbar$, in
quantum mechanics.  In our formalism, `$\LAMBDA$' plays the same role
which is played by the Planck constant, `$\hbar$' in quantum mechanics.
The traditional beam-optics is completely obtained from our formalism
in the limit $\LAMBDA \longrightarrow 0$.

Noting, that $M_z^{- 1} = M_z = \beta$, we multiply both sides of
equation~(\ref{intermediate}) by
\bea
\left[
\ba{cc}
M_z & {\mbox{\boldmath $0$}} \\
{\mbox{\boldmath $0$}} & M_z
\ea
\right]^{- 1}
=
\left[
\ba{cc}
\beta & {\mbox{\boldmath $0$}} \\
{\mbox{\boldmath $0$}} & \beta
\ea
\right]
\eea
and $(\i \LAMBDA )$\,, then, we obtain
\bea
\i \LAMBDA \frac{\partial }{\partial z}
\left[
\ba{cc}
\psi^{+} (\r_\perp , z) \\
\psi^{-} (\r_\perp , z)
\ea
\right]
& = &
\hat{\H}_g
\left[
\ba{cc}
\psi^{+} (\r_\perp , z) \\
\psi^{-} (\r_\perp , z)
\ea
\right]\,.
\label{G-H}
\eea
This is the basic optical equation, where
\bea
\hat{\H}_g
& = &
- n_0
\left[
\ba{cc}
\beta & {\mbox{\boldmath $0$}} \\
{\mbox{\boldmath $0$}} & - \beta
\ea
\right]
+ \hat{\cal E}_g + \hat{\cal O}_g \nn \\
\hat{\cal E}_g
& = &
- \left(n \left(\r \right) - n_0 \right)
\left[
\ba{cc}
\beta & {\mbox{\boldmath $0$}} \\
{\mbox{\boldmath $0$}} & \beta
\ea
\right] \beta_g \nn \\
& & \quad
+
\left[
\ba{ccc}
\beta \left\{
{\mbox{\boldmath $M$}}_\perp  \cdot \p_\perp
-  \i \LAMBDA
{\mbox{\boldmath $\Sigma$}} \cdot {\mbox{\boldmath $u$}}
\right\}
& &
{\mbox{\boldmath $0$}} \nn \\
{\mbox{\boldmath $0$}}
& &
\beta \left\{
{\mbox{\boldmath $M$}}_\perp^{*}  \cdot \p_\perp
-  \i \LAMBDA
{\mbox{\boldmath $\Sigma$}}^{*} \cdot {\mbox{\boldmath $u$}}
\right\}
\ea
\right] \nn \\
\hat{\cal O}_g
& = &
\left[
\ba{ccc}
{\mbox{\boldmath $0$}} & &
- \LAMBDA
\left({\mbox{\boldmath $\Sigma$}} \cdot {\mbox{\boldmath $w$}}\right)
\alpha_y \\
- \LAMBDA
\left({\mbox{\boldmath $\Sigma$}}^{*} \cdot {\mbox{\boldmath $w$}}\right)
\alpha_y
& &
{\mbox{\boldmath $0$}}
\ea
\right]
\label{G-Partition}
\eea
where `$g$' stands for {\em grand}, signifying the eight dimensions
and
\bea
\beta_g
=
\left[
\ba{cc}
{\mbox{\boldmath $I$}} & {\mbox{\boldmath $0$}} \\
{\mbox{\boldmath $0$}} & - {\mbox{\boldmath $I$}}
\ea
\right]\,.
\eea
The above optical Hamiltonian is exact (as exact as the Maxwell
equations in a time-independent linear media).  The approximations
are made only at the time of doing specific calculations.  Apart
from the exactness, the optical Hamiltonian is in complete algebraic
analogy with the Dirac equation with appropriate physical
interpretations.  The relevant point is:
\bea
\beta_g \hat{\cal E}_g
=
\hat{\cal E}_g \beta_g\,, \qquad \quad
\beta_g \hat{\cal O}_g
=
- \hat{\cal O}_g \beta_g\,.
\eea
We note that the upper component ($\Psi^{+}$) is coupled to the lower
component ($\Psi^{-}$) through the  logarithmic divergence of the
resistance function.  If this coupling function,
${\mbox{\boldmath $w$}} = 0$ or is approximated to be zero, then the
equations for ($\Psi^{+}$) and ($\Psi^{-}$) get completely decoupled,
leading to two independent equations.  Each of these two equations is
equivalent to the other.  These are the leading equations for our
studies of beam-optics and polarization.  In the optics context any
contribution from the gradient of the resistance function can be
assumed to be negligible.  With this reasonable assumption we can
decouple the equations and reduce the problem from eight dimensions
to four dimensions.  In the following sections we shall present a
formalism with the approximation ${\mbox{\boldmath $w$}} \approx 0$.
After constructing the formalism in four dimensions we shall also
address the question of dealing with the contributions coming from
the gradient of the resistance function.  This will require the
application of the Foldy-Wouthuysen transformation technique in
{\em cascade} as we shall see.  This justifies the usage of the two
derived laboratory functions in place of permittivity and permeability
respectively.

\section{The Beam-Optical Formalism}
In the previous section, starting with the Maxwell equations we
presented the exact representation of the Maxwell equations
using $8 \times 8$ matrices.  From this representation we constructed
the optical Hamiltonian having $8 \times 8$ matrices.  The coupling of
the upper and lower components of the corresponding eight-vector was
neatly expressed through the logarithmic divergence of the laboratory
function,  the resistance.  We reason that in the optical context we
can safely ignore this term and reduce the problem from eight to four
dimensions without any loss of physical content.

We drop the `$^{+}$' throughout and then the beam-optical Hamiltonian
is
\bea
\i \LAMBDA \frac{\partial }{\partial z} \psi \left(\r \right)
& = &
\hat{\H} \psi \left(\r \right) \nn \\
\hat{\H}
 & = &
- n_0 \beta + \hat{\cal E} + \hat{\cal O} \nn \\
\hat{\cal E}
& = &
- \left(n \left(\r \right) - n_0 \right) \beta
- \i \LAMBDA \beta
{\mbox{\boldmath $\Sigma$}} \cdot {\mbox{\boldmath $u$}} \nn \\
\hat{\cal O}
& = &
\i \left(M_y p_x - M_x p_y \right) \nn \\
& = &
\beta \left({\mbox{\boldmath $M$}}_{\perp} \cdot \hatp_{\perp} \right)\,.
\label{H-Partition}
\eea
If we were to neglect the derivatives of the permittivity and
permeability, we would have missed the term,
$(- \i \LAMBDA \beta
{\mbox{\boldmath $\Sigma$}} \cdot {\mbox{\boldmath $u$}})$.  This
is an outcome of the exact treatment.

Proceeding with our analogy with the Dirac equation: this extra term
is analogous to the anomalous magnetic/electric moment term coupled to
the magnetic/electric field respectively in the Dirac equation.
The term we dropped (while going from the exact to the almost-exact) is
analogous to the anomalous magnetic/electric moment term coupled to the
electric/magnetic fields respectively.  However it should be born in
mind that in our exact treatment, both the terms were derived from the
Maxwell equations, where as in the Dirac theory the anomalous terms
are added based on experimental results and certain arguments of
invariances.  Besides, these are the only two terms one gets.
The term,
$(- \i \LAMBDA \beta
{\mbox{\boldmath $\Sigma$}} \cdot {\mbox{\boldmath $u$}})$
is related to the polarization and we shall call it as the
{\em polarization term}.

One of the other similarities worth noting, relates to the square of
the optical Hamiltonian.
\bea
\hat{\H}^2
& = &
\left\{
n^2 \left(\r \right)
- \hatp_{\perp}^2 \right\} - \LAMBDA^2 u^2
+ \left[
{\mbox{\boldmath $M$}}_{\perp} \cdot \hatp_{\perp} \,,
n \left(\r \right) \right] \nn \\
& & \quad
+ 2 \i \LAMBDA n (\r)
{\mbox{\boldmath $\Sigma$}} \cdot {\mbox{\boldmath $u$}}
+ \i \LAMBDA \left[
{\mbox{\boldmath $M$}}_{\perp} \cdot \hatp_{\perp} \,,
{\mbox{\boldmath $\Sigma$}} \cdot {\mbox{\boldmath $u$}}
\right] \nn \\
& = &
\left\{
n \left(\r \right)
+ \i \LAMBDA
{\mbox{\boldmath $\Sigma$}} \cdot {\mbox{\boldmath $u$}}
\right\}^2 - \hatp_{\perp}^2 \nn \\
& & \quad
+
\left[
{\mbox{\boldmath $M$}}_{\perp} \cdot \hatp_{\perp} \,,
\left\{n \left(\r \right)
+ \i \LAMBDA
{\mbox{\boldmath $\Sigma$}} \cdot {\mbox{\boldmath $u$}}
\right\} \right]
\label{SQUARE}
\eea
It is to be noted that the square of the Hamiltonian in our formalism
differs from the square of the Hamiltonian in the square-root
approaches~\cite{DFW,Dragt-Wave} and the scalar approach
in~\cite{KJS-1,Khan-1}.  This is essentially the same type of
difference which exists in the Dirac case.  There too, the square of
the Dirac Hamiltonian gives rise to extra pieces (such
as, $- \hbar q {\mbox{\boldmath $\Sigma$}} \cdot \B$, the
Pauli term which couples the spin to the magnetic field) which is
absent in the Schr\"{o}dinger and the  Klein-Gordon descriptions.
It is this difference in the square of the Hamiltonians which
give rise to the various extra wavelength-dependent contributions
in our formalism.  These differences persist even in the approximation
when the polarization term is neglected.

Recalling, that in the traditional scalar wave theory for treating
monochromatic quasiparaxial light beam propagating along the positive
$z$-axis, the $z$-evolution of the optical wave function $\psi(\r)$ is
taken to obey the Schr\"{o}dinger-like equation
\bea
\i \LAMBDA\frac{\partial }{\partial z} \psi (\r)
= \hat{H} \psi (\r)\,,
\label{Schr}
\eea
where the optical Hamiltonian $\hat{H}$ is formally given by the
radical
\bea
\hat{H} = - \left({n^2 (\r) - \hatp_\perp^2} \right)^{1/2}\,,
\eea
and $n (\r) = n (x , y , z)$.  In beam optics the rays are assumed to
propagate almost parallel to the optic-axis, chosen to be $z$-axis,
here.  That is, $\left| \hatp_\perp \right| \ll 1$.
The refractive index is the order of unity.
For a medium with uniform refractive index, $n (\r) = n_0$ and
the Taylor expansion of the radical is
\bea
\left({n^2 (\r) - \hatp_\perp^2} \right)^{1/2}
& = &
n_0 \left\{1 - \frac{1}{n_0^2} \hatp_\perp^2 \right\}^{1/2} \nn \\
& = &
n_0 \left\{
1 - \frac{1}{2 n_0^2} \hatp_\perp^2
- \frac{1}{8 n_0^4} \hatp_\perp^4
- \frac{1}{16 n_0^6} \hatp_\perp^6 \right. \nn \\
& & \left. \quad \qquad \qquad
- \frac{5}{128 n_0^8} \hatp_\perp^8
- \frac{7}{256 n_0^{10}} \hatp_\perp^{10} - \cdots
\right\}\,.
\eea
In the above expansion one retains terms to any desired degree of
accuracy in powers of
$\left(\frac{1}{n_0^2} \hatp_\perp^2\right)$.  In general the
refractive index is not a constant and varies.  The variation of the
refractive index $n (\r)$, is expressed as a Taylor expansion in the
spatial variables $x$, $y$ with $z$-dependent coefficients.  To get
the beam optical Hamiltonian one makes the expansion of the radical
as before, and retains terms to the desired order of accuracy in
$\left(\frac{1}{n_0^2} \hatp_\perp^2\right)$ along with all the other
terms (coming from the expansion of the refractive index $n (\r)$) in
the phase-space components up to the same order.  In this expansion
procedure the problem is partitioned into paraxial behaviour $+$
aberrations, order-by-order.

In relativistic quantum mechanics too, one has the problem of
understanding the behaviour in terms of nonrelativistic limit $+$
relativistic corrections, order-by-order.  In the Dirac theory of the
electron this is done most conveniently through the Foldy-Wouthuysen
transformation~\cite{Foldy,BD}.  The Hamiltonian derived
in~(\ref{H-Partition}) has a very close algebraic resemblance with the
Dirac case, accompanied by the analogous physical interpretations.
The details of the analogy and the Foldy-Wouthuysen transformation are
given in Appendix-A.

To the leading order, that is to order,
$\left(\frac{1}{n_0^2} \hatp_\perp^2\right)$ the beam-optical
Hamiltonian in terms of $\E$ and $\O$ is formally given by
\bea
\i \LAMBDA \ddz \left| \psi \right\rangle
& = &
\hat{\cal H}^{(2)} \left|\psi \right\rangle\,, \nn \\
\hat{\cal H}^{(2)} & = &
- n_0 \beta + \E - \frac{1}{2 n_0} \beta \O^2\,.
\label{FW-2-Formal}
\eea
Note that $\O^2 = - \hatp_\perp^2$ and
$\E = - \left(n \left(\r \right) - n_0 \right) \beta
- \i \LAMBDA \beta
{\mbox{\boldmath $\Sigma$}} \cdot {\mbox{\boldmath $u$}}$.
Since, we are primarily interested in the forward propagation,
we drop the $\beta$ from the non-matrix parts of the Hamiltonian.
The matrix terms are related to the polarization.  The formal
Hamiltonian
in~(\ref{FW-2-Formal}), expressed in terms of the phase-space
variables is:
\bea
\hat{\cal H}^{(2)}
& = &
- \left\{n \left(\r \right)
- \frac{1}{2 n_0} \hatp_{\perp}^2 \right\}
- \i \LAMBDA \beta
{\mbox{\boldmath $\Sigma$}} \cdot {\mbox{\boldmath $u$}}\,.
\label{H-Two}
\eea
Note that one retains terms up to quadratic in the Taylor expansion
of the refractive index $n (\r)$ to be consistent with the order
of $\left(\frac{1}{n_0^2} \hatp_\perp^2\right)$.  This is the paraxial
Hamiltonian which also contains an extra matrix dependent term, which
we call as the polarization term.  Rest of it is similar to the one
obtained in the traditional approaches.

To go beyond the paraxial approximation one goes a step further in
the Foldy-Wouthuysen iterative procedure.  Note that, $\O$ is the
order of $\hatp_\perp$.  To order
$\left(\frac{1}{n_0^2} \hatp_\perp^2\right)^2$,
the beam-optical Hamiltonian in terms of $\E$ and $\O$ is formally
given by
\bea
\i \LAMBDA \ddz \left|\psi \right\rangle
& = &
\hat{\cal H}^{(4)} \left|\psi \right\rangle\,, \nn \\
\hat{\cal H}^{(4)}
& = &
- n_0 \beta + \E - \frac{1}{2 n_0} \beta \O^2 \nn \\
& & - \frac{1}{8 n_0^2}
\left[\O ,
\left(\left[\O , \E \right] + \i \LAMBDA \ddz \O \right) \right] \nn \\
& & + \frac{1}{8 n_0^3} \beta
\left\{
 \O^4
+
\left(\left[\O , \E \right] + \i \LAMBDA \ddz \O \right)^2
\right\}\,.
\label{FW-4-Formal}
\eea
Note that $\O^4 = \hatp_\perp^4$, and $\ddz \O = 0$.
The formal Hamiltonian in~(\ref{FW-4-Formal})
when expressed in terms of the phase-space variables is
\bea
\hat{\cal H}^{(4)}
& = &
-
\left\{n (\r)  - \frac{1}{2 n_0} \hatp_\perp^2
- \frac{1}{8 n_0^3} \hatp_\perp^4 \right\} \nn \\
& &
- \frac{1}{8 n_0^2}
\left\{
\left[\hatp_\perp^2 \,, \, \left(n (\r) - n_0 \right) \right]_{+}
 \right. \nn \\
& &
\left. \qquad \qquad \qquad
\vphantom{\hatp_{\perp}^2}
+ 2 \left( p_x \left(n (\r) - n_0 \right) p_x
+
p_y \left(n (\r) - n_0 \right) p_y \right) \right\} \nn \\
& &
- \frac{\i}{8 n_0^2}
\left\{
\left[p_x \,, \, \left[p_y \,, \left(n (\r) - n_0 \right) \right]_{+}
\right]\,
-
\left[p_y \,, \left[p_x \,, \left(n (\r) - n_0 \right) \right]_{+}
\right]
\right\} \nn \\
& &
+ \frac{1}{8 n_0^3}
\left\{
\left[p_x \,, \left(n (\r) - n_0 \right)\right]_{+}^2
+
\left[p_y \,, \left(n (\r) - n_0 \right) \right]_{+}^2
\right\} \nn \\
& &
+ \frac{\i}{8 n_0^3}
\left\{
\left[
\left[p_x \,, \, \left(n (\r) - n_0 \right) \right]_{+} \,, \,
\left[p_y \,, \, \left(n (\r) - n_0 \right) \right]_{+}
\right]
\right\} \nn \\
& & \quad
\cdots
\label{H-Four}
\eea
where $[A , B]_{+} = (AB + BA)$ and `$\cdots$' are the contributions
arising from the presence of the polarization term.  Any further
simplification would require information about the refractive
index $n (\r)$.

Note that, the paraxial Hamiltonian~(\ref{H-Two}) and the leading order
aberration Hamiltonian~(\ref{H-Four}) differs from the ones derived in
the traditional approaches.  These differences arise by the presence of
the wavelength-dependent contributions which occur in two guises.  One
set occurs totally independent of the polarization term in the basic
Hamiltonian.  This set is a multiple of the unit matrix or at most the
matrix $\beta$.  The other set involves the contributions coming from
the polarization term in the starting optical Hamiltonian.  This gives
rise to both matrix contributions and the non-matrix contributions, as
the squares of the polarization matrices is unity.   We shall discuss
the contributions of the polarization to the beam optics elsewhere.
Here, it suffices to note existence of the the wavelength-dependent
contributions in two distinguishable guises, which are not present in
the traditional prescriptions.

\section{When ${\mbox{\boldmath $w$}} \ne 0$}
In the previous sections we assumed,
${\mbox{\boldmath $w$}} = 0$ and this enabled us to develop a
formalism using $4 \times 4$ matrices {\em via} the
Foldy-Wouthuysen machinery.  The Foldy-Wouthuysen transformation
enables us to eliminate the odd part in the $4 \times 4$ matrices,
to any desired order of accuracy.  Here too we have the identical
problem, but a step higher in dimensions.  So, we need to apply the
Foldy-Wouthuysen to reduce the strength of the odd part in eight
dimensions.  This will reduce the problem from eight to four
dimensions.

We start with  the grand optical equation in~(\ref{G-H})
and proceed with the Foldy-Wouthuysen transformations as before,
but with each quantity in double the number of dimensions.
Symbolically this means:
\bea
& &
\hat{\H}
\longrightarrow
\hat{\H}_g\,, \qquad
\psi
\longrightarrow
\psi_g
=
\left[
\ba{cc}
\psi^{+} \\
\psi^{-}
\ea
\right]\,, \nn \\
& &
\hat{\cal E}
\longrightarrow
\hat{\cal E}_g\,, \qquad
\hat{\cal O}
\longrightarrow
\hat{\cal O}_g\, \nn \\
& &
n_0
\longrightarrow
n_g
=
n_0
\left[
\ba{cc}
\beta & {\mbox{\boldmath $0$}} \\
{\mbox{\boldmath $0$}} & - \beta
\ea
\right]\,.
\eea
The first Foldy-Wouthuysen iteration gives
\bea
\hat{\cal H}_g^{(2)}
& = &
- n_0
\left[
\ba{cc}
\beta & {\mbox{\boldmath $0$}} \\
{\mbox{\boldmath $0$}} & - \beta
\ea
\right]
+ \E_g - \frac{1}{2 n_0} \beta_g \O_g^2\, \nn \\
& = &
- n_0
\left[
\ba{cc}
\beta & {\mbox{\boldmath $0$}} \\
{\mbox{\boldmath $0$}} & \beta
\ea
\right] \beta_g
+ \E_g
+ \frac{1}{2 n_0} \LAMBDA^2
{\mbox{\boldmath $w$}} \cdot {\mbox{\boldmath $w$}}
\left[
\ba{cc}
\beta & {\mbox{\boldmath $0$}} \\
{\mbox{\boldmath $0$}} & - \beta
\ea
\right] \beta_g\,.
\label{G-2-Formal}
\eea
We drop the $\beta_g$ as before and then get the following
\bea
\i \LAMBDA \frac{\partial }{\partial z} \psi \left(\r \right)
& = &
\hat{\H} \psi \left(\r \right) \nn \\
\hat{\H}
 & = &
- n_0 \beta + \hat{\cal E} + \hat{\cal O} \nn \\
\hat{\cal E}
& = &
- \left(n \left(\r \right) - n_0 \right) \beta
- \i \LAMBDA \beta
{\mbox{\boldmath $\Sigma$}} \cdot {\mbox{\boldmath $u$}}
+
\frac{1}{2 n_0} \LAMBDA^2 w^2 \beta \nn \\
\hat{\cal O}
& = &
\i \left(M_y p_x - M_x p_y \right) \nn \\
& = &
\beta \left({\mbox{\boldmath $M$}}_{\perp} \cdot \hatp_{\perp} \right)\,,
\label{}
\eea
where,
$w^2 = {\mbox{\boldmath $w$}} \cdot {\mbox{\boldmath $w$}}$, the square
of the logarithmic gradient of the resistance function.  This is how
the basic optical Hamiltonian~(\ref{H-Partition}) gets modified.  The
next degree of accuracy is achieved by going a step further in the
Foldy-Wouthuysen iteration and obtaining the $\hat{\cal H}_g^{(4)}$.
Then, this would be the higher refined starting optical Hamiltonian,
further modifying the basic optical Hamiltonian~(\ref{H-Partition}).
This  way we can apply the Foldy-Wouthuysen in {\em cascade} to obtain
the higher order contributions coming from the logarithmic gradient of
the resistance function, to any desired degree of accuracy.  We are
very unlikely to need any of these contributions, but it is possible
to keep track of them.

\section{Concluding Remarks}
We start with the Maxwell equations and express them in a matrix form
in a medium with varying permittivity and permeability in presence of
sources using $8 \times 8$ matrices.  From this exact matrix
representation we construct the exact optical Hamiltonian for a
monochromatic quasiparaxial light beam.  The optical Hamiltonian has
a very close algebraic similarity with the Dirac equation.  We
exploit this similarity to adopt the standard machinery, namely the
Foldy-Wouthuysen transformation technique of the Dirac theory.  This
enabled us to obtain the beam-optical Hamiltonian to any desired degree
of accuracy.  We further get the wavelength-dependent contributions to
at each order, starting with the lowest-order paraxial paraxial
Hamiltonian.

The beam-optical Hamiltonians also have the wavelength-dependent
matrix terms which are associated with the polarization.  In this
approach we have been able to derive a Hamiltonian which contains both
the beam-optics and the polarization.  In Part-III~\cite{Khan-4} we
shall apply the formalism to the specific examples and see how the
beam-optics (paraxial behaviour and the aberrations) gets modified by
the wavelength-dependent contributions.
In Part-IV~\cite{Khan-5} we shall examine the polarization
component of the formalism presented here.

\renewcommand{\theequation}{A.{\arabic{equation}}}
\setcounter{equation}{0}

\begin{center}

{\Large\bf
Appendix-FW \\
Foldy-Wouthuysen Transformation
} \\

\end{center}

In the traditional scheme the purpose of expanding the {\em light
optics} Hamiltonian
$\hat{H} = - \left(n^2 (\r) - \hatp_\perp^2\right)^{1/2}$ in a
series using $\left(\frac{1}{n_0^2} \hatp_\perp^2\right)$ as the
expansion parameter is to understand the propagation of the
quasiparaxial beam in terms of a series of approximations
(paraxial + nonparaxial).  Similar is the situation in the case of the
{\em charged-particle optics}.  Let us recall that in relativistic
quantum mechanics too one has a similar problem of understanding
the relativistic wave equations as the nonrelativistic approximation
plus the relativistic correction terms in the quasirelativistic regime.
For the Dirac equation (which is first order in time) this is done most
conveniently using the Foldy-Wouthuysen transformation leading to an
terative diagonalization technique.

The main framework of the formalism of optics, used here (and in the
charged-particle optics) is based on the transformation technique of
the Foldy-Wouthuysen theory which casts the Dirac equation in a form
displaying the different interaction terms between the Dirac particle
and and an applied electromagnetic field in a nonrelativistic and
easily interpretable form (see,~\cite{Foldy}-\cite{Acharya}, for a
general discussion of the role of the Foldy-Wouthuysen-type
transformations in
particle interpretation of relativistic wave equations).
In the Foldy-Wouthuysen theory the Dirac equation is decoupled through
a canonical transformation into two two-component equations: one
reduces to the Pauli equation in the nonrelativistic limit and the
other describes the negative-energy states.

Let us describe here briefly the standard Foldy-Wouthuysen theory so
that the way it has been adopted for the purposes of the above studies
in optics will be clear.
Let us consider a charged-particle of rest-mass $m_0$, charge $q$
in the presence of an electromagnetic field characterized by
$\El = - \Nab \phi
- \frac{\partial }{\partial t} {\mbox{\boldmath $A$}}$
and $\B = \Nab \times {\mbox{\boldmath $A$}}$.
Then the Dirac equation is
\bea
\i \hbar \frac{\partial}{\partial t} \Psi(\r , t)
& = & \hat{H}_D \Psi(\r , t)
\label{A-FW-1} \\
\hat{H}_D
& = &
{m_0 c^2} \beta + q \phi + c \Al \cdot \hat{\vpi} \nn \\
& = &
{m_0 c^2} \beta + \hat{\cal E} + \hat{\cal O} \nn \\
\hat{\cal E}
& = &
q \phi \nn \\
\hat{\cal O}
& = &
c \Al \cdot \hat{\vpi}\,,
\label{A-FW-2}
\eea
where
\bea
{\mbox{\boldmath $\alpha$}}
& = &
\left[
\ba{cc}
{\mbox{\boldmath $0$}} & {\mbox{\boldmath $\sigma$}} \nn \\
{\mbox{\boldmath $\sigma$}} & {\mbox{\boldmath $0$}}
\ea
\right]\,, \qquad
\beta
=
\left[
\ba{cc}
\one & {\mbox{\boldmath $0$}} \nn \\
{\mbox{\boldmath $0$}} & - \one
\ea
\right]\,, \qquad
\one
=
\left[
\ba{cc}
1 & 0 \nn \\
0 & 1
\ea
\right]\,, \nn \\
{\mbox{\boldmath $\sigma$}}
& = &
\left[
\sigma_x =
\left[
\ba{cc}
0 & 1 \\
1 & 0
\ea
\right]\,, \
\sigma_y =
\left[
\ba{lr}
0 & - \i \\
\i & 0
\ea
\right]\,, \
\sigma_z =
\left[
\ba{lr}
1 & 0 \\
0 & -1
\ea
\right]
\right].
\eea
with
$
\hat{\vpi}
=
{\hat{\mbox{\boldmath $p$}}} - q {\mbox{\boldmath $A$}}$,
$\hat{\mbox{\boldmath $p$}} = - \i \hbar \Nab$, and
$\hat{\pi}^2 =
\left(\hat{\pi}_x^2 + \hat{\pi}_y^2 + \hat{\pi}_z^2\right)$.

In the nonrelativistic situation the upper pair of components of the
Dirac Spinor $\Psi$ are large compared to the lower pair of components.
The operator $\hat{\cal E}$ which does not couple the large and small
components of $\Psi$ is called `even' and $\hat{\cal O}$ is called an
`odd' operator which couples the large to the small components.
Note that
\beq
\beta \hat{\cal O} = - \hat{\cal O} \beta\,, \qquad
\beta \hat{\cal E} = \hat{\cal E} \beta\,.
\eeq
Now, the search is for a unitary transformation,
$\Psi'$ $=$ $\Psi$ $\longrightarrow$ $\hat{U} \Psi$, such that the equation
for $\Psi '$ does not contain any odd operator.

In the free particle case (with $\phi = 0$ and $\hat{\vpi} = \hat{\p}$)
such a Foldy-Wouthuysen transformation is given by
\bea
\Psi \longrightarrow \Psi' & = & \hat{U}_{F} \Psi \nn \\
\hat{U}_F & = & e^{\i \hat{S}} =
e^{\beta \Al \cdot \hat{\p} \theta} \,,\quad {\rm tan}\,2 |\hat{\p}|
\theta = \frac{| \hat{\p}|}{m_0 c}\,.
\eea
This transformation eliminates the odd part completely from the
free particle Dirac Hamiltonian reducing it to the diagonal form:
\bea
\i \hbar \frac{\partial}{\partial t} \Psi'
& = &
e^{\i \hat{S}} \left({m_0 c^2} \beta + c \Al \cdot \hat{\p} \right)
e^{- \i \hat{S}} \Psi ' \nn \\
& = &
\left(\cos \,| \hat{\p}| \theta +
\frac{\beta \Al \cdot \hat{\p}}{| \hat{\p} |} \sin \,| \hat{\p} | \theta
\right)
\left({m_0 c^2} \beta + c \Al \cdot \hat{\p} \right) \nn \\
& & \qquad \qquad
\times \left(\cos \,| \hat{\p}| \theta -
\frac{\beta \Al \cdot \hat{\p}}{| \hat{\p} |} \sin \,| \hat{\p} | \theta
\right) \Psi' \nn \\
& = &
\left(m_0 c^2 \cos \,2 | \hat{\p}| \theta + c
| \hat{\p} | \sin \,2 | \hat{\p} | \theta \right) \beta \Psi' \nn \\
& = &
\left(\sqrt{m_0^2 c^4 + c^2 \hat{p}^2} \right) \beta \,\Psi'\,.
\eea

In the general case, when the electron is in a time-dependent
electromagnetic field it is not possible to construct an
$\exp (\i \hat{S})$ which removes the odd operators from the transformed
Hamiltonian completely. Therefore, one has to be content with a
nonrelativistic expansion of the transformed Hamiltonian in a
power series in $1/m_0 c^2$ keeping through any desired order.
Note that in the nonrelativistic case, when $|\p | \ll m_0 c$,
the transformation operator $\hat{U}_F = \exp (\i \hat{S})$ with
$\hat{S} \approx - \i \beta \hat{\cal O} /2 m_0 c^2$, where
$\hat{\cal O} = c \Al \cdot \hat{\p} $ is the odd part of the
free Hamiltonian.  So, in the general case we can start with the
transformation
\beq
\Psi^{(1)} = e^{\i \hat{S}_1} \Psi, \qquad \hat{S}_1 =
- \frac{\i \beta \hat{\cal O}}{2 m_0 c^2}
= - \frac{\i \beta \Al \cdot \hat{\vpi }}{2 m_0 c}\,.
\eeq
Then, the equation for $\Psi^{(1)}$ is
\bea
\i \hbar \frac{\partial}{\partial t} \Psi^{(1)}
& = &
\i \hbar \frac{\partial}{\partial t} \left(e^{\i \hat{S}_1}
\Psi \right)
=
\i \hbar \frac{\partial}{\partial t} \left(e^{\i \hat{S}_1}
\right) \Psi + e^{\i \hat{S}_1} \left(\i \hbar
\frac{\partial}{\partial t} \Psi\right) \nn \\
& = &
\left[\i \hbar \frac{\partial}{\partial t}
\left(e^{\i \hat{S}_1} \right) + e^{\i \hat{S}_1}
\hat{H}_D \right] \Psi \nn \\
& = &
\left[ \i \hbar \frac{\partial}{\partial t}
\left(e^{\i \hat{S}_1} \right)
e^{- \i \hat{S}_1}
+ e^{\i \hat{S}_1} \hat{H}_D e^{- \i \hat{S}_1}
\right] \Psi^{(1)} \nn \\
& = &
\left[
e^{\i \hat{S}_1} \hat{H}_D e^{- \i \hat{S}_1}
- \i \hbar e^{\i \hat{S}_1}
\frac{\partial}{\partial t} \left(e^{- \i \hat{S}_1} \right)
\right] \Psi^{(1)} \nn \\
& = &
\hat{H}_D^{(1)} \Psi^{(1)}
\eea
where we have used the identity $\frac{\partial}{\partial t}
\left(e^{ \hat{A}} \right) e^{- \hat{A}}$ $+$ $e^{\hat A}
\frac{\partial}{\partial t} \left(e^{ -\hat{A}} \right)$
$=$ $\frac{\partial}{\partial t} \hat{I}$ $= 0$.

Now, using the identities
\bea
e^{\hat{A}} \hat{B} e^{-\hat{A}}
& = & \hat{B} + [\hat{A} , \hat{B} ]
+ \frac{1}{2!} [\hat{A} , [ \hat{A} , \hat{B} ]]
+ \frac{1}{3!} [\hat{A} , [ \hat{A} , [ \hat{A} , \hat{B} ]]] +
\ldots \nn \\
& & e^{\hat{A}(t)} \frac{\partial}{\partial t} \left(e^{-\hat{A}(t)} \right)
\nn \\
& & \ \ = \left( 1 + {\hat{A}(t)} + \frac{1}{2!} {\hat{A}(t)}^2
+ \frac{1}{3!} {\hat{A}(t)}^3 \cdots \right) \nn \\
& & \ \ \quad \quad \times \frac{\partial}{\partial t}
\left( 1 - {\hat{A}(t)} + \frac{1}{2!} {\hat{A}(t)}^2
- \frac{1}{3!} {\hat{A}(t)}^3 \cdots \right) \nn \\
& & \ \
= \left(1 + \At + \frac{1}{2!} \At^2
+ \frac{1}{3!} \At^3 \cdots \right) \nn \\
& & \ \ \quad \quad
\times \left(- \dAt + \frac{1}{2!} \left\{\dAt \At +
\At \dAt \right\} \right. \nn \\
& & \ \ \quad \quad
- \frac{1}{3!} \left\{\dAt \At^2 + \At \dAt \At \right.  \nn \\
& & \ \ \quad \quad \left. \left.
+ \At^2 \dAt \right\} \ldots \right) \nn \\
& & \ \ \approx
- \dAt - \frac{1}{2!} \left[\At , \dAt \right] \nn \\
& & \ \ \quad \quad
- \frac{1}{3!} \left[\At , \left[\At , \dAt \right] \right] \nn \\
& & \ \ \quad \quad
- \frac{1}{4!} \left[\At , \left[ \At , \left[ \At , \dAt
\right] \right] \right]\,,
\eea
with $\hat{A} = {\i \hat{S}_1}$, we find
%
\bea
\hat{H}_D^{(1)} & \approx & \hat{H}_D - \hbar \dsone
+ \i \left[\sone , \hat{H}_D - \frac{\hbar}{2} \dsone \right] \nn \\
& & \qquad
- \frac{1}{2!} \left[ \sone , \left[\sone ,
\hat{H}_D - \frac{\hbar}{3} \dsone \right] \right] \nn \\
& & \qquad
- \frac{\i}{3!} \left[ \sone , \left[\sone , \left[\sone ,
\hat{H}_D - \frac{\hbar}{4} \dsone \right] \right] \right]\,.
\label{A-FW-9}
\eea
Substituting in~(\ref{A-FW-9}),
$\hat{H}_D = {m_0 c^2} \beta + \hat{\cal E} + \hat{\cal O}$, simplifying the
right hand side using the relations $\beta \hat{\cal O} = - \hat{\cal O} \beta$
and $\beta \hat{\cal E} = \hat{\cal E} \beta$ and collecting everything
together, we have
\bea
\hat{H}_D^{(1)}
& \approx &
{m_0 c^2} \beta + \hat{\cal E}_1 + \hat{\cal O}_1 \nn \\
\hat{\cal E}_1
& \approx &
\E + \frac{1}{2 m_0 c^2} \beta \O^2 - \frac{1}{8 m_0^2 c^4}
\left[\O ,
\left( \left[\O , \E \right] +  \i \hbar \dO \right) \right] \nn \\
& & \quad
- \frac{1}{8 m_0^3 c^6} \beta \O^4 \nn \\
\O_1
& \approx & \frac{\beta}{2 m_0 c^2}
\left(\left[\O , \E \right] + \i \hbar \dO \right)
- \frac{1}{3 m_0^2 c^4} \O^3\,,
\eea
with $\E_1$ and $\O_1$ obeying the relations
$\beta \hat{\cal O}_1 = -
\hat{\cal O}_1 \beta$ and $\beta \hat{\cal E}_1 = \hat{\cal E}_1 \beta$
exactly like $\E$ and $\O$.  It is seen that while the term $\O$ in
$\hat{H}_D$ is of order zero with respect to the expansion parameter
$1/{m_0 c^2}$
({\em i.e.}, $\O$ $=$ $O \left( \left( 1/{m_0 c^2} \right)^0 \right)$
the odd part of
$\hat{H}_D^{(1)} $, namely $\O_1$, contains only terms of
order $1/{m_0 c^2}$ and higher powers of $1/{m_0 c^2}$
({\em i.e.}, $\O_1 = O \left( \left(1/{m_0 c^2}\right) \right)$).

To reduce the strength of the odd terms further in the transformed
Hamiltonian a second Foldy-Wouthuysen transformation is applied with
the same prescription:
\bea
\Psi^{(2)}
& = & e^{\i \hat{S}_2} \Psi^{(1)} \,, \nn \\
\qquad \hat{S}_2
& = &
- \frac{ \i \beta \hat{\cal O}_1}{2 m_0 c^2} \nn \\
& = &
- \frac{\i \beta}{2 m_0 c^2} \left[
\frac{\beta}{2 m_0 c^2}
\left( \left[\O , \E \right] + \i \hbar \dO \right)
- \frac{1}{3 m_0^2 c^4} \O^3 \right]\,. 
\eea
After this transformation,
\bea
\i \hbar \frac{\partial}{\partial t} \Psi^{(2)}
& = & \hat{H}_D^{(2)} \Psi^{(2)}\,, \quad
\hat{H}_D^{(2)}
=
{m_0 c^2} \beta + \hat{\cal E}_2 + \hat{\cal O}_2 \nn \\
\hat{\cal E}_2
& \approx &
\E_1\,, \quad
\O_2 \approx \frac{\beta}{2 m_0 c^2}
\left(\left[\O_1 , \E_1 \right] + \i \hbar
\frac{\partial \O_1}{\partial t} \right)\,, 
\eea
where, now, $\O_2 = O \left(\left(1/{m_0 c^2}\right)^2 \right)$.
After the third transformation
\beq
\Psi^{(3)} = e^{\i \hat{S}_3}\,\Psi^{(2)}, \qquad \hat{S}_3 =
- \frac{ \i \beta \hat{\cal O}_2}{2 m_0 c^2} \nn
\eeq
we have
\bea
\i \hbar \frac{\partial}{\partial t} \Psi^{(3)}
& = &
\hat{H}_D^{(3)} \Psi^{(3)}\,, \quad
\hat{H}_D^{(3)}
=
{m_0 c^2} \beta + \hat{\cal E}_3 + \hat{\cal O}_3 \nn \\
\hat{\cal E}_3 & \approx & \E_2 \approx \E_1\,, \quad
\O_3 \approx \frac{\beta}{2 m_0 c^2} \left(\left[\O_2 , \E_2 \right]
+ \i \hbar \frac{\partial \O_2}{\partial t} \right)\,, 
\eea
where $\O_3 = O \left( \left( 1/{m_0 c^2} \right)^3 \right)$. So,
neglecting $\O_3$,
\bea
\hat{H}_D^{(3)}
& \approx &
{m_0 c^2} \beta + \hat{\cal E} +
\frac{1}{2 m_0 c^2} \beta \hat{\cal O}^2 \nn \\
& & \quad
- \frac{1}{8 m_0^2 c^4} \left[\O , \left( \left[\O , \E \right]
+ \i \hbar \frac{\partial \O }{\partial t} \right) \right] \nn \\
& & \quad
-
\frac{1}{8 m_0^3 c^6} \beta
\left\{
\O^4
+
\left(\left[\O , \E \right] + \i \hbar
\frac{\partial \O }{\partial t} \right)^2
\right\}
\label{A-FW-FOUR}
\eea
It may be noted that starting with the second transformation
successive $(\E , \O)$ pairs can be obtained recursively using the
rule
\bea
\E_j & = & \E_1 \left(\E \rightarrow \E_{j-1} ,
\O \rightarrow \O_{j-1} \right) \nn \\
\O_j
& = &
\O_1 \left(\E \rightarrow \E_{j-1} ,
\O \rightarrow \O_{j-1} \right)\,, \quad j > 1\,,
\eea
and retaining only the relevant terms of desired order at each step.

With $\hat{\cal E} = q \phi$ and
$\hat{\cal O} = c \Al \cdot \hat{\vpi}$, the final reduced
Hamiltonian~(\ref{A-FW-FOUR}) is, to the order calculated,
\bea
\hat{H}_D^{(3)}
& = &
\beta \left({m_0 c^2} + \frac{\hat{\pi}^2}{2 m_0}
- \frac{\hat{p}^4}{8 m_0^3 c^6} \right) + q \phi
- \frac{ q \hbar}{2 m_0 c} \beta \Vsig \cdot \B \nn \\
& & \quad
- \frac{\i q {\hbar}^2}{8 m_0^2 c^2} \Vsig \cdot
{\rm curl}\,{\mbox{\boldmath $E$}}
- \frac{q{\hbar}}{4 m_0^2 c^2} \Vsig \cdot
{\mbox{\boldmath $E$}} \times \hat{\p} \nn \\
& & \quad
- \frac{q{\hbar}^2}{8 m_0^2 c^2}
{\rm div}{\mbox{\boldmath $E$}}\,,
\eea
with the individual terms having direct physical interpretations. The
terms in the first parenthesis result from the expansion of
$\sqrt{m_0^2 c^4 + c^2 \hat{\pi}^2}$
showing the effect of the relativistic mass increase. The second and
third terms are the electrostatic and magnetic dipole energies. The
next two terms, taken together (for hermiticity), contain the
spin-orbit interaction. The last term, the so-called Darwin term,
is attributed to the {\em zitterbewegung} (trembling motion) of the
Dirac particle: because of the rapid coordinate fluctuations over
distances of the order of the Compton wavelength ($2 \pi \hbar /m_0 c$)
the particle sees a somewhat smeared out electric potential.

It is clear that the Foldy-Wouthuysen transformation technique expands
the Dirac Hamiltonian as a power series in the parameter
$1/{m_0 c^2}$ enabling the use of a systematic approximation
procedure for studying the deviations from the nonrelativistic
situation.  We note the analogy between the nonrelativistic
particle dynamics and paraxial optics:

\begin{center}
{\bf The Analogy}
\end{center}
\begin{tabular}{ll}
{\bf Standard Dirac Equation} ~~~~~~~ & {\bf Beam Optical Form} \\
$m_0 c^2 \beta + \E_D + \O_D$ & $- n_0 \beta + \E + \O$ \\
$m_0 c^2$ & $- n_0$ \\
Positive Energy & Forward Propagation \\
Nonrelativistic, $\left|\mbox{\boldmath $\pi$}\right| \ll m_0 c$ &
Paraxial Beam, $\left|\hatp_\perp \right| \ll n_0$ \\
Non relativistic Motion  & Paraxial Behavior \\
~~ + Relativistic Corrections & ~~ + Aberration Corrections \\
\end{tabular}

\bigskip

Noting the above analogy, the idea of Foldy-Wouthuysen form of the
Dirac theory has been adopted to study the paraxial optics and
deviations from it by first casting the Maxwell equations in a spinor
form resembling exactly the Dirac equation~(\ref{A-FW-1}, \ref{A-FW-2})
in all respects: {\em i.e}., a multicomponent $\Psi$ having the upper
half of its components large compared to the lower components and the
Hamiltonian having an even part $(\E)$, an odd part $(\O)$, a suitable
expansion parameter, ($|\hatp_\perp|/{n_0} \ll 1$) characterizing the
dominant forward
propagation and a leading term with a $\beta$ coefficient commuting
with $\E$ and anticommuting with $\O$.  The additional feature of
our formalism is to return finally to the original representation
after making an extra approximation, dropping $\beta$ from the final
reduced optical Hamiltonian, taking into account the fact that we are
primarily interested only in the forward-propagating beam.

\end{document}